%% file: sample-sigconf-authordraft.tex
\documentclass[sigconf]{acmart}
\usepackage{booktabs}
\usepackage{soul}       
\usepackage{color}
\usepackage{array}
\usepackage{makecell}
\usepackage{tcolorbox}
\usepackage{fontawesome5}  

\tcbuselibrary{skins, breakable}

\newtcolorbox{Implication}{
  enhanced,
  breakable,
  colback=yellow!8!white,
  colframe=orange!70!black,
  fonttitle=\bfseries\small,
  title={$\blacktriangleright$ Critical Implication},
  attach boxed title to top left={yshift=-2mm, xshift=4mm},
  boxed title style={
    colback=orange!70!black,
    colframe=orange!70!black,
    sharp corners
  },
  sharp corners,
  left=6pt,
  right=6pt,
  top=8pt,
  bottom=6pt,
  boxrule=0.6pt,
  fontupper=\small
}
\AtBeginDocument{%
  }

\begin{document}

\title{Essential, Yet Overlooked: Identity Verification Barriers for Blind and Low Vision People in Government Services}

\author{Ryan John Oommen}
\email{rjo5309@psu.edu}
\affiliation{%
  \institution{Penn State}
  \city{State College}
  \state{PA}
  \country{USA}
}

\author{Tanusree Sharma}
\email{tanusree.sharma@psu.edu}
\affiliation{%
  \institution{Penn State}
  \city{State College}
  \state{PA}
  \country{USA}
}








\begin{abstract}
 Identity verification is a critical gateway to accessing government services and public benefits, yet contemporary systems are typically designed around visual interaction, leaving blind and low vision (BLV) individuals disproportionately burdened. In this work, we examine how BLV users navigate identity verification in government services and how current designs shape their access, security, and autonomy. Through a mixed-methods study combining analysis of 219 Reddit posts and semi-structured interviews with 16 BLV participants, we uncover systemic accessibility breakdowns across both digital and in-person verification processes. Our findings show that inaccessible verification workflows do not merely inconvenience users—they restructure how security is achieved in practice. We also identify how repeated verification demands, inaccessible physical infrastructure, and policy changes exacerbate exclusion from essential services. At the same time, participants articulate complex perspectives on AI, viewing it as both a critical accessibility aid and a growing vector for identity fraud. 
 
\end{abstract}



\begin{teaserfigure}
  \includegraphics[width=\textwidth]{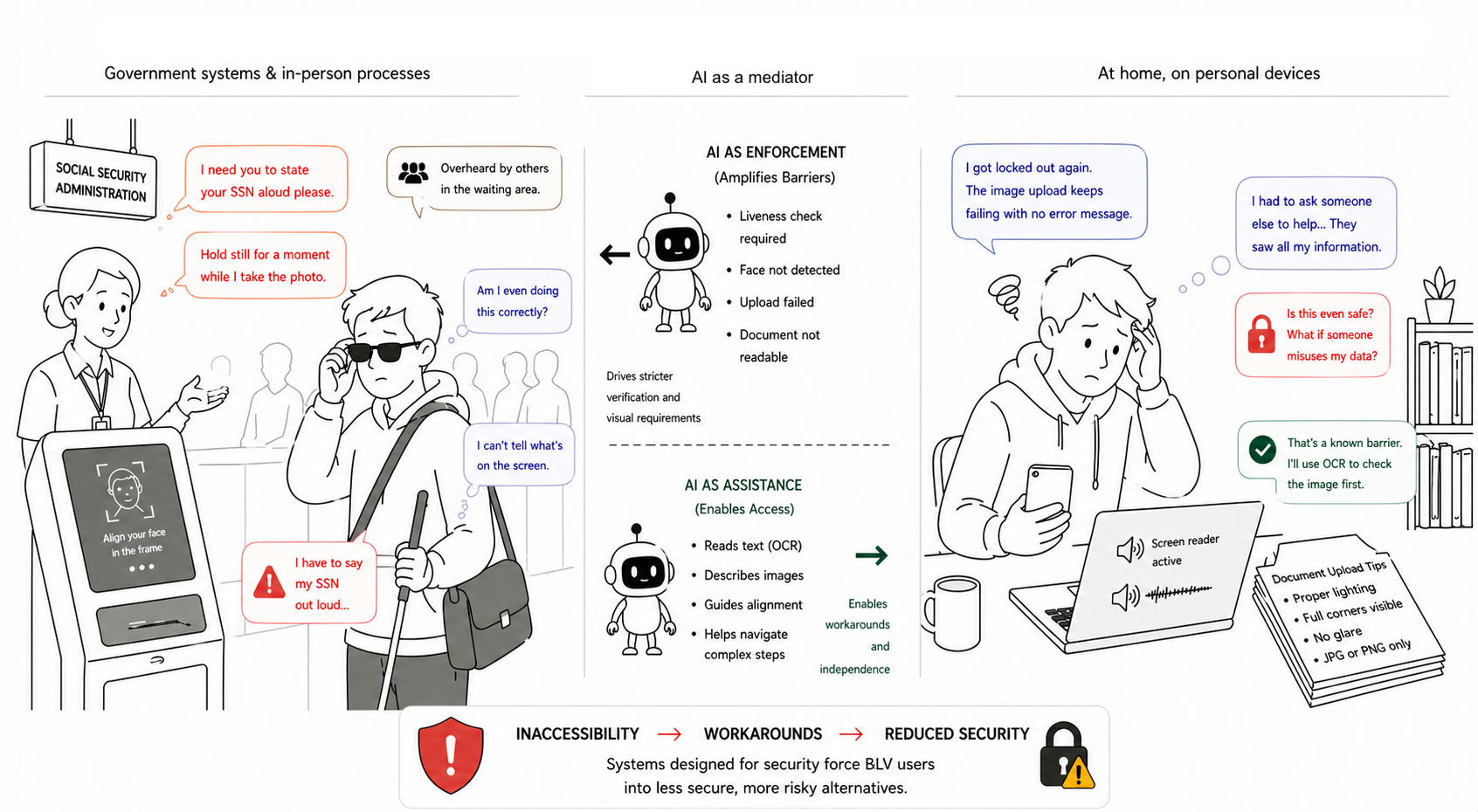}
  \caption{Figure 1 presents an illustration of identity verification workflows experienced by BLV individuals. The left panel depicts in-person verification challenges, including inaccessible kiosks, reliance on others, and forced verbal disclosure of sensitive information such as Social Security numbers. The right panel shows at-home verification struggles, such as repeated login failures, inaccessible interfaces, and dependence on others for assistance. As as a mediator show dual role that may introduce barriers (e.g., liveness checks, unreadable documents) and as an assistive tool that enables access.}
  \Description{Figure 1 presents a conceptual illustration of identity verification workflows experienced by blind and low vision (BLV) individuals. The left panel depicts in-person verification challenges, including inaccessible kiosks, reliance on others, and forced verbal disclosure of sensitive information such as Social Security numbers. The right panel shows at-home verification struggles, such as repeated login failures, inaccessible interfaces, and dependence on others for assistance. AI as a mediator shows dual role that may introduce barriers (e.g., liveness checks, unreadable documents) and as an assistive tool that enables access.}
  \label{fig:teaser}
\end{teaserfigure}


\maketitle
\input{section/Introduction}
\input{section/Related-Work}
\input{section/Method}
\input{section/Results}
\input{section/Discussion}

\bibliographystyle{ACM-Reference-Format}
\bibliography{sample-base}

\end{document}

%% file: section/Introduction.tex
\section{Introduction}
Identity verification is a fundamental gateway to accessing digital services and public benefits. Services, such as, Social Security Disability Insurance (SSDI), Supplemental Security Income (SSI), Medicare, other federal programs rely on processes that requires them to prove who they are claiming to be. This is typically done through document-based checks, biometric authentication, or a mix of online and in-person verification. However, the threat landscape around these systems is becoming increasingly complex. Advances in artificial intelligence, especially deepfakes and synthetic identity have significantly heightened fraud risks, including benefit theft and identity impersonation. In response, agencies and service providers have strengthened verification mechanisms to better distinguish legitimate users from automated or malicious actors~\cite{ssa2025blog, gao2025retirement}. However, these increasingly stringent systems often assume visual interaction as a baseline as reflected in many public infrastructures~\cite{li2023toucha11y}. 

For the millions of Americans who are blind or have low vision, common verification steps, such as selfie-based facial recognition, document photo capture, image-based CAPTCHAs and emerging form of verification, such as liveness (e.g. ID.me) can be difficult to complete independently~\cite{stanley2022facialrecognition}. These challenges are often compounded by the format of verification (e.g., physical versus digital) and the rapid introduction of new technologies. As a result, measures designed to improve security can simultaneously create significant barriers to access, effectively limiting equitable participation in essential government services. While the security community has explored decades to hardening authentication systems against adversarial attack~\cite{grassi2017digital}, the usability and accessibility~\cite{akter2023if, kamarushi2022onebuttonpin, saxena2009authentication, lazar2012soundsright} of high tech systems, comparatively little research on examining how people with visual impairment navigate government benefit verification, often nearlt absent from literature. Most studies treat authentication barriers for BLV users as a usability inconvenience rather than a civil rights and safety issue~\cite{toussaint2025inclusive, saxena2009authentication, zhao2018face, zhao2018face}. Whereas, studies on government fraud prevention treat the beneficiary population as a homogeneous public rather than a diverse group that includes people for whom standard verification tools are effectively inaccessible or coercive~\cite{sorrell2023digitalauthentication, nysfocus2023idme}.

To address this gap, in this work, we examined three research questions:

\textbf{RQ1:} What are the experience of blind and low vision individuals in identity verification system for government services? 

\textbf{RQ2:} How does the current identity verification system for government services create barriers or risks for blind and low vision users?

\textbf{RQ3:} How do blind and low vision users perceive the role of AI in threatening identity systems?

To answer these three research questions, we employed a mixed-methods approach. We analyzed \textbf{219 Reddit posts} followed by a semi-structured interview with \textbf{16 blind and blow vision individuals} to c
identify the risks and barrier with current identity verification in government services and to explore the strategies they currently employ to mediate and mitigate potential harms. Then, we explored their perception regarding the role of AI in both threatening and potentially enabling improvement for identity systems.


Our \textbf{contributions }are threefold:

(a)  First, we surface a set of previously underdocumented accessibility and security experiences of BLV users within the specific, high-stakes context of government benefit identity verification, a context largely absent from the security and accessibility literature. Our work highlights persistent barriers in both digital and physical workflows, including inaccessible biometric checks, document capture failures, and in-person systems that require public disclosure of sensitive information. These constraints frequently force users into dependency on others or alternative channels that introduce new security and privacy risks.

(b) We further show that emerging AI-driven identity threats intensify this tension. While institutions adopt stricter verification techniques to counter synthetic identity fraud, these same techniques often deepen accessibility gaps. Participants, however, demonstrate adaptive strategies and articulate clear expectations for more inclusive systems.

(c) We argue that current identity verification infrastructures reflect a systemic misalignment between security policy and accessibility needs to support multiple verification pathways, accessible system feedback, and verification models that do not rely solely on visual or biometric assumptions. Rather than treating accessibility as a secondary concern, we position it as central to equitable and effective security design. 


Our work contributes to timely policy discussions. In early 2025, the Social Security Administration, under pressure from the Department of Government Efficiency, tightened identity proofing by removing phone-based verification for many services and requiring either online verification through my Social Security which depends on biometrics and document uploads often inaccessible to blind and low-vision (BLV) users or in-person visits, even as field offices were closing nationwide~\cite{kimery2025ssa}. Disability advocates and lawmakers argued that these changes disproportionately harmed elderly, rural, and disabled populations~\cite{zippel2025ssa}. Our findings provide concrete direction towards toward policy as a central force for structural improvement in government services. 

%% file: section/Related-Work.tex
\section{Related Work}
Our study sits at the intersection of literature in (a) accessible authentication research for BLV users; (b) identity verification in government benefit contexts; and (3) AI-driven identity fraud. 

\subsection{Authentication \& Identity Verification Challenges for Blind and Low Vision Users}
Blind and low vision users face challenges in authenticating to digital services have been documented since at least 2007, when Holman et.al first systematically investigated the security-related obstacles blind users encounter on the web~\cite{holman2007developing}. This work established that standard security mechanisms, particularly CAPTCHA, were frequently inaccessible which forces users into workarounds that introduced new risks. This observation initiated substantial line of research on CAPTCHA accessibility~\cite{lazar2012soundsright}.
Empirical study~\cite{dosono2015m}
of in-situ observation on authentication routines across computers, smartphones, and websites showed that BLV users suffered significant delays, frequently encountered inaccessible error messages, and were forced to develop elaborate workarounds when login workflows failed. Some of these workaround, such as storing passwords in plaintext files compatible with screen readers introduced security vulnerabilities not faced by sighted users~\cite{dosono2018toward}. This line of work  argued explicitly for \textit{"accessible authentication"} as a design goal for the security community. 
A large-scale studies with 325 BLV users across twelve countries about their mobile authentication preferences and strategies found that most participants used familiar personal information (names, dates) to construct memorable passwords, a practice that trades security for memorability and that fingerprint biometrics were rated as the most accessible and trusted authentication method, while PINs were rated the least secure~\cite{briotto2018understanding}. This underscored that these are not idiosyncratic problems of a particular platform or region but structural failures in how authentication systems are designed.
More recently, study shows significant accessibility gaps where randomly generated passwords could not be read aloud by screen readers, warning messages appeared as unlabeled popups, and many participants fell back to reusing simple passwords or maintaining Braille format password lists that posed their own physical security risks~\cite{ponticello2025blind}. This confirm again that \textit{"accessibility is not a luxury but a basic prerequisite for digital security"} - a framing directly relevant to our work on government benefit verification.  

\subsection{Privacy, Security, and Cooperative Coping Among BLV Users}
Beyond authentication mechanics, research shows that blind and low-vision (BLV) users often rely on trusted allies, including, family, caregivers, or volunteers in privacy and security-sensitive tasks that standard tools do not support independently~\cite{hayes2019cooperative}. This \textit{“cooperative security”} model is practically necessary but exposes sensitive credentials, health, and financial data to third parties~\cite{sharma2024m}.
Subsequently, studies reinforces these risks where BLV users report awareness of threats such as surveillance, accidental disclosure, and caregiver exploitation, yet remain underserved by mainstream privacy tools~\cite{ahmed2015privacy}. They also express discomfort with camera-based assistive technologies that may inadvertently capture~\cite{akter2020uncomfortable, sharma2025before, zhang2024designing, sharma2023disability}. These concerns can be heightened in government benefit systems requiring biometric and document submission. A missing opportunity in prior work is the accessibility of physical verification infrastructure. Many government identity systems still require BLV users to appear in person for initial enrollment or periodic re-verification under strict deadline. As documented in accessibility and public service reports, such barriers often lead disabled individuals to rely on third parties or miss deadlines, increasing both exclusion and exposure to fraud or coercion~\cite{acharya2025accessibility, davidson2024accessibility, npr2026digitalaccessibility}.
These failures are not only usability issues but security risks: when systems exclude BLV users from independently completing required steps, they push users toward insecure workarounds. This reinforces the argument that accessibility must be treated as a core security property rather than a secondary concern~\cite{wang2018inclusive}.


\subsection{Identity Verification in Government Benefit Contexts}
Despite the extensive literature on digital authentication accessibility, research specifically examining identity verification in government services is sparse. Government services, such as, Social Security, SSDI, SSI, Medicaid, SNAP, housing assistance involve high-stakes, often compulsory verification workflows where failure to authenticate successfully can mean loss of income, healthcare, or housing. Yet the HCI and security communities have largely left this domain understudied. One of the few systematic analyses is Digital Government Hub’s 2022–2024 landscape study, which maps authentication and identity-proofing requirements across major public benefits programs~\cite{sorrell2023digitalauthentication}. It highlights a core tension: fraud prevention introduces friction, while accessibility and equity require minimizing it. The report identifies biometric facial comparison (selfie matched to ID) as particularly problematic, creating barriers for users who cannot complete visual capture tasks and raising concerns about data retention and privacy.

The SSA's own evolution of identity proofing policy has been largely instructive where agency's 2021 transition to Login.gov and ID.me for digital identity verification introduced facial recognition and document upload workflows. In early 2025, under pressure from DOGE, the SSA announced a major tightening of identity proofing requirements. Many users who could not complete online verification through the \textit{"my Social Security"} portal which relies on selfie matching and document upload would be required to verify in person at a field office~\cite{kimery2025ssa}. This policy was announced concurrently with the closure of dozens of field offices and reductions in SSA workforce, creating a situation where neither the online nor the in-person pathway was reliably accessible to vulnerable populations. After significant pushback from disability advocates, lawmakers, and the AARP, the SSA partially walked back the policy, exempting SSDI, SSI, and Medicare applications from the in-person requirement, but the episode crystallized the absence of disability-inclusive impact assessment in SSA security policy~\cite{npr2025ssa_identity}.
The Center on Budget and Policy Priorities has documented that over 10\% of seniors in 35 U.S. states would need to travel more than 45 miles to reach their closest SSA field office, and that roughly 8 million older Americans have a medical condition or disability that makes travel physically difficult~\cite{zippel2025ssa}. When identity verification policy eliminates remote options without accessible digital alternatives, the resulting system is one that disproportionately excludes precisely the populations that programs like SSDI and SSI exist to serve. 

For BLV users, who are among the largest groups receiving these benefits, this represents a systemic exclusion.
Accessibility barriers in government services remain largely unaddressed and, which our study focus on investigating.

\subsection{AI-Driven Fraud and the Emergence of Personhood Credentials}
The policy pressure driving the SSA's verification tightening is not merely bureaucratic. Generative AI has produced a qualitative shift in the fraud landscape. AI systems can now generate photorealistic images of nonexistent people, animate still photos into convincing videos, clone voices from short audio samples, and produce spoofed selfie images of synthetic identities holding fabricated government documents~\cite{adler2024personhood}. In 2023, deepfake incidents in the fintech sector increased 700\% year-over-year. Deloitte's Center for Financial Services projects that AI-enabled fraud losses in the United States could reach \$40 billion by 2027, up from \$12.3 billion in 2023~\cite{grant2026fightaiidentityfraud}. Globally, identity fraud losses exceeded \$50 billion in 2025~\cite{fintechglobal2026deepfakes}. 

The canonical technical response was requiring stronger biometric verification, liveness detection, and document authenticity checks~\cite{zhou2023iterative, sharma2025verifying}. This also creates the greatest accessibility barriers for BLV users. When the SSA mandates facial liveness checks to defeat deepfake spoofing, it simultaneously creates a workflow that a user relying on a screen reader cannot independently complete. This is not a coincidence or an edge case; it is a structural property of visual biometric verification design. The security hardening that the fraud environment demands and the accessibility requirements of the BLV population are, under current design paradigms, in direct tension.


%% file: section/Method.tex
\section{Study Methodology}
To answer research questions about how current identity verification system for government services (e.g., Social Security) create barriers or risks for blind and low vision users, we conducted a content analysis of social media
(Reddit), and semi-structured interviews with 16 BLV individuals.

\subsection{Reddit Data}
\textbf{Collection.} To obtain the understanding of public sentiment and discussions around the Social Security verification landscape, we conducted a computational analysis of Reddit data. We selected Reddit as the primary data source due to its large library of detailed and experience-driven narratives related to government services and personal challenges. Reddit also allowed for access to unfiltered accounts of system interaction and personal anecdotes. The Python Reddit API Wrapper (PRAW) scanned Reddit for posts matching a set of keywords. We compiled the results in a structured Excel format for further analysis. 
We initiated the search with a set of search queries that included both identity verification specific terms and broad terms related to government services. Phrases such as \textit{"identity verification," “social security fraud,” “SSI fraud,” “SSDI fraud,” “disability fraud,” “social security overpayment,” “social security delay.”} We executed queries through multiple subreddits containing topics related to Government and Social Security verification. Data was collected from\textit{r/socialsecurity, r/SSDI, r/Disability, r/personalfinance, and r/legaladvice}. The dataset contained a total of 1,091 rows. After manual review, we identified 219 posts that are related to security issues in the context of blindness and low vision. 




\textbf{Analysis.} We applied thematic and inductive analysis to organize our data to gain a clear understanding. Two researchers independently read 20\% of the reddit data and developed codes, and compared them until arriving at a consistent codebook. Then, first author coded the rest of the data. 
We followed an open coding and deductive analysis method to explore participants’ practices, challenges, and design suggestions of identity verification in government services. 
We derived 43 low level code which we organized into high level categories for BLV users: \textbf{Inaccessible verification process in govt. services (34 post), Login \& authentication failures (52 posts), Identity Impersonation \& Fraud  (91 post), Distrust in current govt. service resulted from identity verification issue (13 pasts), Screen reader /
CAPTCHA barriers (4 posts), AI \& emarging threat in identity verification (3 posts) and Administrative delays
\& errors (22 posts). }
The insights gained from this Reddit analysis served as a foundation for the interview phase and the identification of key challenges.

\%end{table}

\subsection{User Interviews}
Interview study allowed us to better understand user experiences with identity verification in government services. The objective of this study was to gain understanding of user perceptions with the current system while gauging user expectation of possible solution. We recruited a total of 13 participants using participant mailing lists. This study is approved by IRB. We conducted an hour interview with each participants. 

\textbf{Interview Procedure.} All interviews were conducted remotely through Zoom being audio recorded and transcribed with participant consent. 
We first with began with a brief introduction outlining the purpose and format of the study. Interviews were structured into four main sections. In the first section, we explored participants current expeirence with identity verification system broadly. Here, participants were asked to describe their understanding of identity verification and their overall background with methods such as one-time passwords and knowledge-based authentication. In second section, we moved into the verification processes in government services to tailor towards the interactions with identity verification in government services. For instance, we asked \textit{``Can you briefly explain about some government related services or  benefits that you use where you were required to verify your identity during enrollment? Could you walk me through the process of how you first enrolled in the benefit program?''} 

We then asked about their experience with fraud related issue in government services. We then dive into understanding their perception towards AI in identity impersonation and fraud that happens in government services. We also asked them to share experience on any emerging identity verification mechanisms they used that provide security against evolving AI related identity risks. For instance \textit{``With AI now capable of mimicking voices and generating realistic facial images or videos, how do you view its potential impact on current identity verification processes? ''}. Finally, we asked their preferences and design recommendation, where participants were asked to propose improvements to current and future systems in response to challenges and concerns they had previously identified.

\textbf{Data Analysis.}
Data collection then ensued by transcribing all recordings and compiling answers into a unified dataset after the completion of all interviews. We applied a similar thematic and deductive approach as we did for Reddit data was used for the interview data.  We began independently reading through transcripts and coding the first round of interview transcripts (4 interviews). We discussed regularly, developing themes and refining them, until a consists themes developed. Weekly, we discussed the emerging themes, resolved any differences, and reached consensus on the interpretations of findings. This iterative process ensured reliable coding framework~\cite{mcdonald2019reliability}. Then, the first author coded the remaining of interview transcripts and developed high-level themes. 


Across the reddit and interview theme, we found Seven thematic challenge areas which overlaps. For each theme, we first report the Reddit-level signal (frequency counts, dominant discourse patterns)~\ref{tab:triangulation}. From this overall frequency, we found AI related identity risks has difference in response where reddit discussion did not have much discusison , however participants' lived experience reflected on this concerns from interview results. 
We present the interview-level elaboration with particular attention to BLV participants' experiences in the subsequent sections. 

\begin{table*}[ht]
\centering
\caption{Cross-Source Triangulation of Challenge Areas in Identity Verfication in Govt Application/Services}
\label{tab:triangulation}
\setlength{\tabcolsep}{8pt}
\renewcommand{\arraystretch}{1.4}
\begin{tabular}{>{\raggedright\arraybackslash}p{2.8cm}
                >{\centering\arraybackslash}p{2cm}
                >{\centering\arraybackslash}p{2cm}
                >{\raggedright\arraybackslash}p{5.5cm}}
\toprule
\textbf{Overlapped Themes}
  & \textbf{\makecell{Reddit\\(N=219)}}
  & \textbf{\makecell{Interviews\\(N=13)}}
  & \textbf{BLV-Specific Signal} \\
\midrule
Inaccessible verification process
  & \makecell{34 posts\\(17\%)}
  & \makecell{9 participants\\(69\%)}
  & CAPTCHA acrobatics; screen reader failures; non-blind-friendly Real ID \\
Login \& authentication failures
  & \makecell{52 posts\\(26\%)}
  & \makecell{13 participants\\(100\%)}
  & Audio CAPTCHA broken; OTP timing inaccessible; error messages unreadable \\
Identity Impersonation \& Fraud  (personal/known)
  & \makecell{91 posts\\(46\%)}
  & \makecell{11 participants\\(84\%)}
  & Dependence on sighted helpers exposes credentials; cooperative security risks \\
Distrust in current system
  & \makecell{13 posts\\(7\%)}
  & \makecell{10 participants\\(76\%)}
  & Verification feels performative; system not designed for disability \\
Screen reader / CAPTCHA barriers
  & \makecell{4 posts\\(2\%)}
  & \makecell{5 participants\\(38\%)}
  & Qualitative signal disproportionate to Reddit count; primary BLV barrier \\
AI \& emerging threat concerns
  & \makecell{3 posts\\(2\%)}
  & \makecell{10 participants\\(76\%)}
  & Voice cloning threatens phone-based workaround BLV users rely on most \\
Administrative delays \& errors
  & \makecell{22 posts\\(11\%)}
  & \makecell{5 participants\\(38\%)}
  & Inaccessible error messages and correction loops amplify burden for BLV users \\
\bottomrule
\end{tabular}
\smallskip
\begin{minipage}{\linewidth}
\end{minipage}
\end{table*}


\subsection{Participant Demographic}
Our study included a total of 16 participants (P1–P16), age ranging from 35 to 65+ years, with the largest subgroup falling within the 35–44 age range, followed by 45–54, 55–64, and 65+ categories. 10 of them were female, 5 male participants and one non-binary participant. Majority of participants were totally blind (n = 11), while the remaining five participants reported some light perception. Most participants (n = 14) reported having government services, including Social Security (SS) services whereas two participants indicated no prior use. 

\begin{table*}[h]
\centering
\caption{Participant Demographics and Background}
\label{tab:demographics}
\begin{tabular}{l l l l l p{8cm}}
\toprule
\textbf{ID} & \textbf{Age} & \textbf{Gender} & \textbf{Blindness} & \textbf{GovtSer} & \textbf{Verification Concerns} \\
\midrule
P1  & 55--64 & Female       & Some Light Perception  & Yes & Authorization issues \\
P2  & 55--64 & Female        & Totally blind & Yes & Identity verification process not secure or accessible; unclear next steps when logging in; confusing page wording and delays \\
P3  & 35--44 & Female        & Some Light Perception & Yes & Frustrating to navigate with a screen reader \\
P4  & 45--54 & Female     & Totally blind   & No  & No challenges reported \\
P5  & 35--44 & Female    &Totally blind     & Yes & CAPTCHA audio options not working; difficulty reading printed/large print forms due to vision problems \\
P6  & 35--44 & Female     &Totally blind    & Yes & Concerned that users without assistance cannot complete identity verification at all \\
P7  & 65+    & Female     &Some Light Perception    & Yes & Cannot independently fill out forms \\
P8  & 45--54 & Male    &Totally blind       & No  & No personal experience; aware others have struggled with identity verification \\
P9  & 35--44 & Male &Totally blind          & Yes & Paper communications should be electronic (web portal/email) for screen reader accessibility \\
P10 & 45--54 & Female  &Some Light Perception       & Yes & New login system (Real ID) is not accessible or blind-friendly; unable to complete registration \\
P11 & 35--44 & non-binary &Totally blind  & Yes & Registration was difficult but the SSA system itself works well for accessibility once set up \\
P12 & 35--44 & Female  &Totally blind       & Yes & No challenges reported \\
P13 & 55--64 & Female &Totally blind        & Yes & Uncertainty about what using the system means; used it for benefit information only, not yet applied \\
P14 & 65+ & Male  &Totally blind       & Yes & Needed help recently for a state ID, and generally find online forms more hindered by usability than accessibility when using a screen reader. \\
P15 & 65+  &Male   & Some Light Perception    & Yes & Delays getting in-person help and frequent inaccessibility of online forms and PDFs \\
P16 & 35--44 & Male  &   Totally blind   & Yes & Assistive technologies like JAWS and refreshable braille displays creates persistent accessibility barriers on government services\\
\bottomrule
\end{tabular}
\end{table*}

%% file: section/Results.tex

\section{Experiences of BLV Individuals with Identity Verification Systems for Government Services (RQ1)}
Participants uniformly described identity verification for government services as a multi-layered challenge that begins at the moment of contact with the system and compounds with each subsequent interaction. Four dominant sub-themes emerged: inaccessible interfaces, forced reliance on others, repetitive verification burdens, and the emotional toll of ongoing friction.

\subsection{Digital Workflows Designed for Sighted User}
Participant encountered design feature in a government verification system that was structurally incompatible with their assistive technology. The most frequently cited failure was the one-time password (OTP) workflow, which required users to leave the verification page to retrieve an emailed code causing the browser session to expire and forcing them to restart the entire process. Whereas, P2 reported that after completing login on the Social Security Administration (SSA) website, 
the \textbf{absence of accessible feedback}
compounds the cognitive load of an already multi-step process. As she stated-
\begin{quote}
\textit{``You put in your information, ask for the one-time password, then go into your Gmail. When you go back, you have to put it all in again and it sends you a new code. So you just keep going in a loop.'' -P1}

\textit{"Every month I sit there wondering whether I got it right or not because once you activate submit or login, there's no completion statement or success message. That's confusing." -P2}
\end{quote}

P16 explained \textbf{document capture requirements} which is central to both login.gov and ID.me, the SSA's mandated third-party identity proofing services represent the\textbf{ single-point failure}. The task of photographing a government-issued ID with precise alignment is visually demanding in ways that standard assistive technology cannot fully mediate. P16 described being entirely locked out of her SSA account because every attempted upload failed the platform's automated image quality check, the system offered no accessible error explanation and no alternative pathway.
P14 elaborates \textbf{compensatory strategies by photographing his ID} at elbow height to stabilize the camera, then passing the captured image through an OCR tool to verify it showed the correct side and adequate text before submitting.
\begin{quote}
\textit{"What I do is try to align the camera as best as I can. After I take the photo, I run it through OCR or AI to make sure it's capturing the correct side with the identifying information."-P14}
\end{quote}
 This \textbf{workaround layering AI assistive technology onto a process} designed for sighted users — enabled some degree of independent completion, but introduced additional steps and points of failure the platform's designers did not anticipate. As P14 explained. Finally, P3 explained challenges regarding \textbf{signature fields within digital verification workflows} presented a distinct barrier where he needed to disable her screen reader entirely to locate and complete a signature field, sign without any visual or tactile feedback about placement accuracy, and re-enable the screen reader. He explained frequent failure to register and required repetition, sometimes escalated to a sighted stranger via a live assistance application. 

From \textbf{Reddit}, 34 posts (17\% of dataset) documented Identity Verification \& Login Issues. One user wrote: \textit{``I can build a computer. I can write code. I have been doing these things since the early 90s... Social Security website is mandating that I verify my identity. I have tried repeatedly over the course of a week. I gave up.''} Another post described the SSA login system as \textit{``circular and backwards''} due to its dependence on requiring other forms of identification. Login \& authentication failures were the second most discussed theme on Reddit (52 posts, 26\%), confirming that OTP loops and document upload failures are not specific to BLV users but are disproportionately experienced by them.

\subsection{Physical Verification Workflows as a Compounding Barrier}
A limitation of prior accessible authentication research is its near-exclusive focus on digital interfaces. Our data reveal that for government identity verification specifically, \textbf{physical workflows} impose barriers of comparable severity. these barriers interacting with digital ones often multiply the overall burden on participants.

Participants described \textbf{in-person verification at SSA} field offices as structured around visual interaction with unassisted kiosks on the office terminals. 
As P3 described the in-person SSA experience succinctly:

\begin{quote}
\textit{"You have to go to a cubicle and put in your information. I have to bring a friend, and I have to say my social security out loud because I can't see the screen. The buttons aren't raised, there's no braille, nothing to help you feel it out. There's a part that says if you're blind, click here, but it doesn't change anything."
— P3}
\end{quote}
The physical inaccessibility of these terminals is not a minor design oversight, it structurally mandates that BLV users be accompanied by a sighted assistant. P9 explained paper-based communication presents a serious physical channel vulnerability. P9 described incident where SSA mail large envelopes containing her full SSN and personal information repeatedly, with documents too large for her mailbox, left in shared building areas visible to neighbors and passersby. She described it not be an incidental privacy risk, a physical data exposure vector built into the \textbf{agency's standard communication protocol}. 
Similarly, P16 shared concerns on document transportation where he had to \textbf{carry original identity documents}, passport, proof of address, medical records on public transit to multiple SSA appointments and acutely aware that loss or theft and he would have had limited means to prevent. This risk is structurally imposed by a system that requires repeated in-person document presentation rather than a verified-once digital credential.

\textbf{Reddit} posts document the SSA's 2025 field office closures noted that 45 SSA office leases were cancelled including five permanent closures in Georgia alone, with users describing drives of 45+ minutes to the nearest office. Another post reported a user who \textit{``went into the office 12 times which is a 45 min drive each time''} over five months still unable to resolve a direct deposit change and identity reverification.
Physical access barriers were corroborated at scale with Reddit's SSA Operations \& Policy category (13 posts) including those who cannot complete online verification into a shrinking in-person infrastructure.

\subsection{Repetitive Reverification as Sustained Burden}
Another theme was frequenly discussed with which verification was required, extending the burden from a one-time enrollment cost to an ongoing, recurring demand. SSA web portal access required fresh authentication on every login. P11 discussed medical benefits systems like Ascension required participants to re-upload identity documents at each appointment booking. P9 described being required to present documents and answer verification questions on every call to the SSA, on every in-person visit, and even after her case had been approved. threat-adaptive policy.
\begin{quote}
"Even after being approved, I had to go back again and repeat the process. For medicare-related things, every time I go into the service, I have to go through all those steps again."-P9
\end{quote}
These reverification burden was not proportional to risk, it appeared to reflect system design defaults rather than just a user experience inconvenience.
It is an occasion that requires mobilizing a sighted helper, disclosing personal information to that helper, and navigating an inaccessible interface under time pressure.

\textbf{Reddit} users described near-identical patterns of administrative reverification loops \textit{``If I filled this out, why would I have to do it again?''} wrote one user required to resubmit already completed paperwork multiple times. A separate post documented a beneficiary caught in a Medicaid and Social Security miscommunication loop \textit{``They claimed they didn't have last year's Medicare denial, even though it was obviously processed since Medicaid was approved. So, I had to fax the Medicare denial letter again, twice.''}
Administrative delays \& errors accounted for 22 Reddit posts (11\%), with resubmission loops appearing as a recurring structural failure independent of any single user's technology or disability status.

\subsection{Emotional and Psychological Toll}Participants articulated the affective dimension of these experiences with consistent clarity. The emotional stakes were not merely frustration with slow systems; they were bound up with dignity, autonomy, and the meaning of disability. As P1 stated \textit{``It's not just the time, it's the frustration because it's a reminder of something I can no longer do by myself.''} P7 framed repeated system failures as producing cumulative trust erosion.
P16 described being locked out of SSA account notifications as producing a material harm such as, loss of access to benefit communications which compounded over time and was not correctable through any channel available to him without sighted help. They charecterize these as civil access failures with ongoing consequences, rather a merely usability problems. 

\textbf{Reddit} posts reflected the same cumulative distress at scale. One user denied disability benefits despite severe health conditions wrote \textit{``They denied me as a Christmas gift.''} Another, marked erroneously as deceased, described their situation as \textit{``sick, ailing, and penniless''} after benefits were immediately terminated. A third user described SSA leadership as \textit{``out of touch,''} arguing that policies were not aligned with current beneficiary realities, echoed widely across the SSA Operations category.
Distrust in the current system appeared in 13 Reddit posts (7\%) highly consistent in tone which reflect systemic rather than incidental grievance.

\section{Barriers and Risks for BLV Users in Identity Verification Systems (RQ2)}
Beyond experiential friction, participants surfaced structural features of current government identity verification systems that produce concrete accessibility barriers and security risks. We term this the \textbf{accessibility-to-security} failure pathway: when a designed channel is inaccessible, users do not abandon the task; they adopt workarounds. Those workarounds systematically undermine the security properties the original channel was designed to protect.  

\subsection{Forced Verbal Disclosure: Accessibility Failure as Privacy Violation}
The most direct instantiation of the accessibility-to-security failure pathway was forced verbal disclosure of sensitive identifiers. P10's description of in-person SSA verification captures the mechanism precisely. She explaind that the kiosk keyboard has no tactile markers, and the only way to complete the task is for a sighted companion to read the screen aloud and for the BLV user to dictate their responses  including their full SSN within earshot of others in the waiting room. 
In this interaction pathways, the security system designed to protect that SSN has structured an interaction that broadcasts it. 

\begin{quote}
\textit{``"I have to say my social security out loud because I can't see the screen... Even in person, there's no accessibility. There's a part that says if you're blind, click here, but it doesn't change anything. It's the same process. There's no solution."— P10}
\end{quote}
P7 described the same exposure in the context of paper form completion - SSA forms that she could not fill out independently required assistance from staff or volunteers she did not know well, who were thereby exposed to her SSN, address, and identifying information.
In the same line, P2 described being required to speak her pharmacy account information aloud at an inaccessible Walmart kiosk. She also discussed that the technology to make these kiosks accessible exist, but it is simply not being deployed.

\textbf{Reddit} posts corroborated verbal and physical PII exposure as a systemic risk rather than an individual failure. One described having their SSN appear on the dark web after the AT\&T breach \textit{``I had 8 alerts about my social security number being on the dark web... I was told to call one of the credit bureaus and put a fraud alert.''} Another documented a case where a user's SSN was used by another person for employment and tax filing, discovered only when the user attempted to file for disability.
Scam, phishing, and breach attempts targeting beneficiaries appeared across the Identity Fraud Accusations \& Investigations category (59 posts, 30\%), confirming that the sensitivity of PII exposed during inaccessible interactions translates directly into real fraud outcomes.


\subsection{Insecure Workarounds Induced by Inaccessible Design}
Beyond verbal disclosure, participants described a second class of accessibility-driven security failures where workaround practices reduce security as a direct response to inaccessible interaction.

Previous literature highlighted knowledge-based authentication (KBA) security questions,  mother's maiden name, first pet, city of birth are already recognized as a weak authentication factor~\cite{ahmed2015privacy} because the answers are often discoverable through public records. As workaround, P1 described deliberately providing false answers to KBA questions and memorizing the fabrications, because correct answers could be found in obituaries or public databases. However, doing so put another participant P6 into risk of locked out of her SSA account and follow a longer process to reverify of uploading new document and biometric-matching with having no recourse that she could complete independently.

\begin{quote}
\textit{"I've tried multiple times to upload pictures of my ID and haven't been successful, so I can't access my benefit letters or notifications anymore." -P6}
\end{quote}

This kind of system lockout has a security dimension that extends beyond inaccessibility. When online access is denied, the participants is pushed toward telephone and mail channels that participants described as less verifiable, more susceptible to interception, and, in the context of the SSA's 2025 policy changes, increasingly unavailable. P1 described a friend whose benefit routing number was altered by fraud.
\begin{quote}
\textit{``Someone got property tax benefits under her name. She kept going through the verification process, but someone intercepted it and changed the routing numbers. She didn't receive the money and only caught it because she reviewed and printed her form''}
\end{quote}

The paper channel, in this case forced on participant as workaround which is itself a fraud vector.

\subsection{Structural Trust Erosion and Anxiety}
Participants' trust in identity systems was substantially shaped by direct experience with data breaches and identity theft. For instance, P2 had been notified of breaches through the state of Texas, Chase Bank, and AT\&T while P3's social security number was compromised when her purse was stolen, and she has maintained a credit freeze ever since. P7 had experienced attempted phone fraud, a relative's identity theft, and a friend's hacked account leading her to stop answering phone calls from unknown numbers entirely.
\begin{quote}
\textit{"I don't trust that anything is 'secure'. I've seen too many systems where data isn't encrypted properly. Do I think it's secure? No. Do I have a choice? No. Even years later, I don't know if someone still has my social security number." -P7}
\end{quote}

Many recognized the systems were insecure but perceived that there are no viable alternative if they wished to access government services. This structural coercion and the absence of meaningful choice were a source of ongoing distress. This aligns with the civil rights reading of accessible authentication proposed by Saxena and Watt~\cite{saxena2009authentication} and reinforced by Toussaint et al.~\cite{toussaint2025inclusive} where the question is not whether participants are technically capable of navigating these systems with assistance, but whether they are entitled to navigate them with equivalent security and independence as sighted users.

\textbf{Reddit's} SSA Operations \& Policy category captured the same structural distrust at the policy level. One post documenting the 2025 DOGE-driven verification tightening noted: \textit{"Tighter identity verification rules to prevent fraud create access barriers for vulnerable recipients."} A cybersecurity professional posting in the same category acknowledged both sides: \textit{"I can appreciate that [the new requirements] can be frustrating at times. However, fraud happens every day"}. This explicitly articulate the security-accessibility tension.

\section{Perceptions of AI's Role in Threatening Identity Systems (RQ3)}

Participants held nuanced and often paradoxical views on AI's role in identity verification. Many perceived AI simultaneously as a tool of empowerment, particularly for assistive and navigational tasks and as an emerging and serious threat to the integrity of identity verification systems. This dual perception structured how participants reasoned about both the promise and risks of AI  in government identity infrastructure.

\subsection{AI as Accessibility Infrastructure}
Several participants noted that AI-based tools, particularly OCR, image description, and voice-assisted navigation had meaningfully expanded their ability to navigate verification processes that would otherwise be entirely inaccessible. P9's use of AI to verify that he had correctly photographed his ID exemplifies this pattern: AI served as a compensatory layer for a gap in system accessibility. As P3 stated \textit{``AI has been a huge benefit, especially for blind people. It has made me more independent and given me more freedom and efficiency in everyday tasks. But in the wrong hands, it can be harmful and intrusive. It could be manipulated to work against me instead of helping me.''}
This framing of AI as assistive infrastructure made the threat dimension more complex. The same technology enabling independence was also, in their view, enabling new forms of identity compromise.

\subsection{Voice and Visual Cloning: AI-Driven Impersonation as Lived Threat}
Participants' threat perceptions were not abstract. Several described direct experience with AI-assisted fraud attempts, and all were aware of the capacity of generative AI to synthesize voices and faces indistinguishable from authentic ones. 
P8 described her boyfriend being targeted by a sophisticated social engineering attack that used personal information, coordinated phone and text messaging, and emotional manipulation to extract money via Apple Pay. The attackers deployed information aggregated from data breaches to simulate legitimacy. P7 also described having heard AI-replicated voices and being unable to distinguish them from authentic ones. This examplifies direct implications for telephone-based verification, which several participants relied on as their primary accessible channel for SSA interaction.

\begin{quote}
\textit{``AI can impersonate people very well. I have heard it do that and you can't really tell the difference. If they can figure out a way to use that for fraud, they probably will. If voice is no longer a reliable authenticator and  visual biometrics are inaccessible to us and then for us voice verification may be precisely the channel most vulnerable to AI-driven attack.''}
\end{quote}

Another incident where P1 described AI-generated voice impersonation targeting older adults in her social network through fake family emergency calls. A neighbor lost \$15,000 to such a scheme. P1's family later establish personal code words with family members to authenticate genuine distress calls - an informal humanness verification protocol developed independently in response to AI threat, without institutional support or guidance.

\subsection{Biometrics as Contested Ground \& Near Potential Solutions}
Participants' dominant technical response to AI-driven identity fraud was stronger biometric verification, visual liveness detection, selfie-to-ID matching. They perceived this as a structural collision,  a hardening that simultaneously increases security against AI impersonation and increases the accessibility barrier for them. 
As P2 said \textit{``We are probably going to have to use biometrics, but AI can also reproduce those, so where is the security?''}

P14 drew a distinction between biometric implementations based on the quality of non-visual guidance they provided. Apple's Face ID, which uses audio prompts and haptic feedback to guide positioning, was navigable independently. ID.me's facial recognition workflow, which relies on visual positioning cues with no non-visual equivalent, required sighted assistance. The technology is the same modality, face-based biometric verification, but the accessibility outcome is entirely determined by whether the implementation includes non-visual guidance. 

\begin{quote}
\textit{``Systems like ID.me are not very accessible for blind users, while systems like Apple Face ID are much better because they use audio and haptic feedback to guide positioning.''}
\end{quote}

Several participants expressed a preference for fingerprint-based verification as a tactile modality that could be completed without visual interaction. P7 and P16 both described fingerprint as their preferred biometric, contingent on consistent recognition accuracy and the assurance that they could complete the process independently. P2 raised a legal dimension that complicated simple biometric preference \textit{``courts have ruled that users can be compelled to unlock devices biometrically but not via password''} - a constraint that shifts the power calculus of biometric adoption in ways that give sophisticated users pause.

Participants also raised concerns about the AI-vulnerability of biometrics themselves. P3 noted that deepfake technology had rendered visual biometrics such as, face recognition, potentially iris scanning, theoretically spoof-able. P14 expected to a privacy-preserving -- \textit{``I would like systems similar to Apple's secure enclave, where biometric data is stored locally and only a secure token is shared instead of raw data.''}

\subsection{Design for AI-Resilient \& Accessible Verification}
Participants articulated specific design demands in response to AI threats and accessibility failures. 

\textbf{Multiple equivalent verification pathways} emerged as the most consistent expectation. P2 articulated that no single verification modality should be mandatory, because every modality has a population for which it is inaccessible. If fingerprint is the only non-visual biometric option and a user lacks fingerprints, no alternative exists. If OTP is the only second factor and SMS autofill is unavailable to a screen reader user on the platform, the session fails. Designing for mandatory single-modality verification is designing for exclusion. 

\begin{quote}
\textit{``We need multiple ways to verify identity. Not everyone can use the same method. Security and accessibility should not be traded off. Both need to be included from the beginning.''— P2}

\end{quote}

\textbf{Human-assisted fallback as a right, not an exception} P16 described positive experience with a company that offered video-call verification with a trained agent were able to complete an identity verification task fully independently and securely, because the human agent could provide real-time guidance for document positioning and confirmation. She noted that no such option existed within SSA or login.gov systems, despite these being the very platforms from which she was currently locked out.

\begin{quote}
\textit{``One time, a company allowed me to verify over a video call where I could show my credentials to a person, and that worked well, but that option is almost never available.''
— P16}

\end{quote}

\textbf{Accessible feedback states throughout verification workflows} Participants expected for feedback workflow accessible throughout, not just at task completion but at each intermediate step. P2 and P11 described uncertainty at multiple points in SSA and login.gov verification flows as a fundamental barrier to independent completion. Accessible feedback is not a supplemental feature; it is a prerequisite for informed interaction.

\textbf{Reusable digital credentials}. Participants described a form of persistent, digital credentials that they can reusable rather than repeated document submission. P14's proposal for a verified-once digital ID stored locally with secure token exchange represents one architectural instantiation of this demand. More broadly, participants were articulating a principle where the verification burden should scale with the actual security risk of the transaction, not default to maximum friction at every interaction point regardless of context or prior verification history.

%% file: section/Discussion.tex
\section{Discussion}
Our findings surface a fundamental design contradiction in the contemporary government identity verification where the mechanisms deployed to harden systems against fraud are, by construction, the mechanisms most inaccessible to blind and low vision users. It is a structural outcome of designing identity systems within a security-first paradigm that has never meaningfully incorporated disability as a design constraint~\cite{renaud2022accessible}. We situate our findings within three bodies of literature, accessible authentication research, privacy and cooperative security among BLV populations, and the emerging AI fraud landscape resulted from identity verification failure.

\subsection{From Accessibility Inconvenience to Civil Rights Failure}
Accessible authentication literature since Holman et al.~\cite{holman2007developing}, that standard security mechanisms create barriers for BLV users to CAPTCHA inaccessibility~\cite{holman2007developing, dosono2015m}, screen-reader-incompatible authentication flows~\cite{dosono2015m}, and password management failures~\cite{ponticello2025blind} have each been characterized as usability problems requiring design improvement. Accessibility as usability framing has produced incremental discussion where CAPTCHA audio alternatives remain structurally inaccessible to screen reader users~\cite{lazar2012soundsright}. Same failure modes Dosono et al.~\cite{davidson2024accessibility} and Ponticello et al.~\cite{ponticello2025blind} show that the field has made minimal progress. We argue that usability problems invite design iteration, they do not compel structural change. 
This specific context of this work on government benefit identity verification is that these are not usability problems. They are civil rights failures with material consequences. Participants' inability to a photo of her ID to login.gov and is thereby locked out of her SSA account, losing access to notifications about the disability benefits on which they depends, verbalizing Social Security Number in a waiting room because the kiosk is visually inaccessible  - are not inconveniences. They are harms of a kind that sighted users of the same system do not experience.

Saxena and Watt~\cite{saxena2009authentication} argued in 2009 that accessible authentication should be understood as a security right rather than a usability accommodation. Then Wang's framework of inclusive security~\cite{wang2018inclusive} extended this argument, contending that security systems which exclude populations on the basis of disability are not merely inaccessible but are actively insecure for those populations, because exclusion forces less secure workarounds. The most recent work by Toussaint et al. on the ALIAS project~\cite{toussaint2025inclusive} represents the most recent attempt to operationalize inclusive security as a design goal for authentication systems. Yet none of these frameworks has been integrated into the design of the government services and verification systems our participants navigate. SSA's 2021 transition to login.gov and ID.me introduced facial recognition and document-upload workflows without apparent disability-inclusive impact assessment~\cite{npr2025ssa_identity, npr2026digitalaccessibility}. The 2025 tightening of SSA identity proofing requirements under DOGE pressure eliminated phone-based verification — the most accessible channel available to BLV users — without providing an accessible digital alternative~\cite{zippel2025ssa}. Our results highlighted that 
the consideration of accessible authentication as solely an usability concern, rather than a policy mandate, has produced seventeen years of limited structural change in day to day applications and services that blind and low vision people leverage and directly impact their livelihood.

\begin{Implication}
Accessible identity verification literature needs reframing its' argument from 
\textit{`these systems are hard to use'} to \textbf{`these systems produce measurable civil rights harms.'} This provides HCI community an opportunity to explicitly position accessible government verification as a compliance matter, not merely a design best-practice.
\end{Implication}

\subsection{Accessibility-to-Security Failure Pathway}
An important theoretical contribution of our findings is a precise account of the mechanism by which accessibility failures produce security failures. This is even more critical when a designed verification channel is inaccessible to a BLV user, that user does not abandon the verification task the task is typically mandatory to access essential services. In our findings, we found participants to adopt the\textbf{ least-bad available alternative} including verbal disclosure of PII to a sighted assistant, delegating device control to a Be My Eyes volunteer, maintaining handwritten credential notes compatible with braille displays, providing deliberately false KBA answers which are systematically less secure than the designed channel. 

This interaction pathways was implicit in Wang~\cite{wang2018inclusive, zhang2023imageally, zhang2021webally} and partially explained in Hayes et al.'s  characterization of cooperative security among BLV users~\cite{hayes2019cooperative} where BLV users routinely share credentials and device access with trusted allies to complete privacy-sensitive tasks that standard tools cannot support independently. While practically necessary, this exposes sensitive data to third parties. Our findings extend this analysis to government benefit verification, where the cooperative security model is not merely an informal coping strategy but the only available path for BLV users whom the formal system has excluded. We also extend it physically where the cooperative security literature has focused on digital credential sharing, but our participants' accounts of in-person SSA kiosks with no tactile markers, paper correspondence with full SSNs left in shared mailrooms, and documents transported on public transit reveal that the pathway operates with equal force in physical verification infrastructure. 

A particularly acute instance of these pathway is the false knowledge-based authentication (KBA) response strategy described by some participants who deliberately provided incorrect answers to security questions, fabricating a pet name, a childhood city to protect against the reality that correct answers are often findable through obituaries, social media, and public records. This is an independently derived user-level response to a known weakness in KBA design that the security literature has documented for over a decade~\cite{grassi2017digital}. While its not irrational behavior, it creates its own vulnerability if the user forgets which fabrications they provided, they are locked out. The system has transferred the security burden entirely onto the user, without providing accessible tools for managing that burden. 

\begin{Implication}
Security researchers and system designers must explicitly model the accessibility-to-security failure pathway when evaluating the security properties of identity verification systems. A system that achieves high theoretical security for a sighted user population while producing low actual security for BLV users through forced workarounds is not a secure system. Security evaluation frameworks that do not account for \textbf{differential accessibility} are measuring security for a subset of users.
\end{Implication}

\subsection{Invisible Accessibility Problem: Physical Verification Infrastructure}
A significant gap in the accessible authentication is the near-exclusive focus on digital interfaces~\cite{dosono2015m, kaushik2023guardlens, wang2018inclusive, ponticello2025blind, feng2024understanding, kamarushi2022onebuttonpin, bhole2024haptic2f} through the lens of digital systems such as, websites, mobile apps, password managers. Even the most recent work~\cite{li2023toucha11y} examine inaccessible public touchscreens which focuses on the touchscreen as a digital interface rather than on the broader physical infrastructure of verification. Yet our participants' reveal that the physical dimensions of government identity verification impose barriers of comparable severity to digital ones and that these barriers are, if anything, less visible  generate since those tangibly do not highlight error logs, accessibility audit trails, or compatibility reports.

In-person SSA field office experience described by multiple participants kiosk keyboards with no tactile markers, no private accommodation for verbal PII disclosure — represents a physical interface failure. The difference is that the digital failure might be caught by an accessibility audit conducted under Section 508 guidelines, however the physical failure is subject to limited systematic review. Davidson~\cite{davidson2024accessibility} and the Digital Government Hub's landscape study~\cite{gao2025retirement} both note that physical verification infrastructure is largely absent from accessible technology policy. Our findings highlights those comparatively less visible risk when transporting original identity documents on public transit to appointments, and to carry printed correspondence containing their full PII through shared residential spaces. These are not hypothetical risks since many of those resulted in participants' benefit payments being diverted because a fraudster intercepted and altered her paper-based routing information during precisely such a physical verification interaction.

Furthermore, policy compounds this physical infrastructure problem where SSA's 2025 decision to eliminate phone-based verification while simultaneously reducing field office staff and closing offices~\cite{ssa2025blog} effectively closed both of the accessible channels BLV users most depended on  telephone and in-person human assistance without providing an accessible digital replacement. Zippel and O'Connor~\cite{zippel2025ssa} document that over 10\% of seniors in 35 states would need to travel more than 45 miles to reach their nearest SSA field office, and that approximately 8 million older Americans have medical conditions making such travel difficult. Our participant directly experienced persistent delays in obtaining in-person help as offices reduced staff while demand from users who could not complete online verification increased. Often this changing policy produced a verification dead end for users who could not complete online proofing and could not access in-person offices. 

\begin{Implication}
Accessible authentication research could be extended its scope to physical verification infrastructure. This includes kiosk interface design, private accommodation spaces for verbal PII disclosure, accessible mail and paper communication formats, and the logistical risks of document transport. 

\end{Implication}

\subsection{AI Threat: Hardening Against Fraud Excludes BLV Users}
Generative AI fraud landscape has produced a genuine and quantitatively serious threat to identity verification. Deepfake incidents in fintech increased 700\% year-over-year in 2023~\cite{grant2026fightaiidentityfraud}. Deloitte's projections of \$40 billion in AI-enabled fraud losses in the United States by 2027~\cite{IC3_PSA_2024_AI_Fraud}. Recent work~\cite{adler2024personhood} highlighted the risks of  generative AI in producing photorealistic images of nonexistent people, animate still photos into convincing videos, clone voices from short audio samples, and generate spoofed selfies of synthetic identities holding fabricated government documents. In response, government moving towards stronger biometric verification, liveness detection, document authenticity checks~\cite{ametefe2024enhancing}. 

Visual liveness detection requires the user to position their face within a camera frame using visual feedback while selfie-to-ID biometric matching requires the user to photograph their own face. Document authenticity checks require the user to photograph their identity documents with precise alignment. Every one of these steps assumes a user who can see what they are doing. Participants broadly preferred biometric modalities over knowledge-based or OTP-based verification, citing independence and reduced cognitive load. First, participants discussed biometric accessibility as implementation-specific, not modality-specific design. The same modality such as, face-based biometric verification produces radically different accessibility outcomes depending on whether the implementation includes non-visual guidance. 

\begin{Implication}
 Liveness detection, selfie biometric, and document-upload workflows should not be deployed in government services without simultaneously providing accessible non-visual equivalents. Biometric authentication standards for government identity verification should specify non-visual guidance as a mandatory accessibility requirement, not an optional enhancement. 

\end{Implication}

\section{Conclusion}
We examined how blind and low vision individuals experience identity verification for government services. Through Reddit analysis and semi-structured interviews with 16 BLV participants, we documented an accessibility-to-security failure pathway in which inaccessible verification design structurally routes BLV users into less secure channels. These failures are not confined to digital interfaces rather physical verification infrastructure and emerging AI fraud landscape exacerbates the stakes. We situated our findings with inclusive security framework and argue that accessibility in identity verification is not a usability accommodation — it is a security property and a civil rights requirement. Until government benefit systems treat BLV users as primary stakeholders rather than edge cases, every tightening of identity verification in the name of security will deepen the exclusion of the people those systems exist to serve.